\documentstyle[psfig]{mn2e}
\newdimen\hssize
\newdimen\hdsize
\begin{document}
\title[Star clusters in NGC 6240]{On the formation of star clusters in the merger NGC 6240}
\author[Pasquali, de Grijs and Gallagher]{A. Pasquali$^{1}$\thanks{E-mail: apasqual@eso.org,
R.deGrijs@sheffield.ac.uk, jsg@astro.wisc.edu}, R. de Grijs$^{2,3}$ and J.S. Gallagher$^{4}$\\
$^{1}$ ESO/ST-ECF, Karl-Schwarzschild-Strasse 2, D-85748 Garching bei M\"unchen, Germany\\
$^{2}$ Institute of Astronomy, University of Cambridge, Madingley Road, Cambridge CB3 0HA \\
$^{3}$ Department of Physics \& Astronomy, University of Sheffield, Hicks Building, Hounsfield
Road, Sheffield S3 7RH\\
$^{4}$ University of Wisconsin, Department of Astronomy, 475 N. Charter St., Madison WI 53706, USA}
\date{}
\pubyear{2003}
\maketitle

\begin{abstract}
We identified star clusters in archived HST/WFPC2 images of the
merger and ultra-luminous infrared galaxy NGC 6240, with the aim of investigating whether
star cluster properties (luminosity, age and mass) in such an extreme
environment differ from those of clusters in less 
luminous starburst galaxies.  We found 54 star clusters 
in all of the F450W, F547M and F814W
exposures, of which 41 are located in the main body of NGC 6240 and 13
in the galactic tails.  Given that only two colours are available to
derive two independent variables (cluster reddening and age), 
we adopted an {\it ad
hoc} procedure to statistically derive cluster parameters under the assumption
that the cluster metallicity is LMC-like.
The colours of each cluster are
fitted to STARBURST99 models of fixed mass and variable ages and reddenings.
All cluster reddening and age solutions with $\chi^2 < 1$
are considered to be consistent with the data.  Masses are derived by
scaling the luminosity of the models with best-fit $\chi^2 < 1$ by the
observed $V$ luminosity, after correction for reddening and distance. 
Therefore, each cluster is described by a range of reddening values,
ages and masses; for each of these parameters we derive probability
functions.  We thus infer that the most probable age of the observed clusters is 
between 5 and 13 Myr and their most probable mass is about $(1 -
2) \times 10^5$ M$_{\odot}$.  
A low probability exists for clusters as
massive as 10$^8$ M$_{\odot}$, as well as for the trend that the mean
cluster mass increases towards the double nuclei of NGC 6240. 
Comparison with star clusters in starburst galaxies seems to point to
more massive clusters being formed in more massive galaxies and 
gas-rich mergers,
while the overall cluster mass distribution might be 
relatively independent of the details of the associated starburst
where dense, massive clusters preferentially form. 
\end{abstract}


\section{Introduction}

The {\it Costar} generation instruments on board 
{\it Hubble Space Telescope} (HST) have significantly
contributed to our understanding of starburst galaxies.  The high
angular resolution of the HST resolved R136a in 30 Doradus,
previously believed to be a single, hyper-massive star (cf.  Moffat
1982; Savage et al. 1983), 
into individual stars (Malumuth \& Heap 1994, Hunter et al. 
1995).  Subsequently, WFPC2 has been used to detect starburst sites and
star-forming regions in nearby and distant galaxies.  UV spectroscopy
with STIS has been performed to study the young stellar content of these
objects, and, in the case of heavy extinction, their identification and
analysis have been carried out with NICMOS (Bushouse et al.  2002,
Farrah et al.  2001 Scoville et al.  2000). 

The class of starbursts includes a variety of galaxy types: spiral
galaxies, blue compact galaxies, interacting galaxies, merging galaxies
and ultra-luminous infrared galaxies (ULIGs) which suggest that the star
formation is triggered by gravitational interactions, mergers and/or bar
formation.  These mechanisms compress the gas which dissipates energy
and moves inward igniting intense star formation (Ashman \& Zepf 2001, Bekki
et al.  2002).  The star formation in starbursts extends to 
high stellar masses ($M_* \simeq 100$ M$_{\odot}$ in R136, Massey
\& Hunter 1998) and is generally characterised by a normal upper initial
mass function (IMF), irrespective of the galaxy metallicity (Leitherer
2000, Gonz\'alez Delgado et al.  2002).  It is estimated that starbursts
produce one quarter of the massive stars in the 
present day Universe (Heckman 1997). 

A common feature of all these galaxies is the formation of super star
clusters (SSCs) with remarkably similar properties, i.e.  age, mass and
luminosity.  The definition of an SSC is based on their compact size, 
young ages of $<$1~Gyr and associated high luminosities, implying
integrated masses typically ranging between 10$^5$ and 10$^6$ M$_{\odot}$.
The upper mass range for SSCs is comparable to the most massive
globular cluster in the Milky Way, $\omega$ Cen.  SSCs are believed to
be the progenitors of globular clusters (cf. Ashman 2002, de Grijs et al. 2003). 

The SSCs in the Antennae are perhaps the best studied so far.  Their
mass is as high as few 10$^6$ M$_{\odot}$ (Zhang \& Fall 1999) 
and they subdivide into four
age groups: the youngest ones ($<$ 5 Myr), located at the edge of the
dust overlap region between the two interacting galaxies, the $5 - 10$ Myr-old 
clusters in the Western loop,
the intermediate-age clusters (100 Myr) found in the NE star-forming
region and the oldest ones ($\simeq$ 500 Myr) in the NW extension of the
Antennae (Whitmore et al.  1999, Zhang et al.  2001).  If the dust
overlap 
region of the two merging galactic nuclei is to be considered the centre
of the Antennae, one could infer that a radial gradient exists in the
mean cluster age, with the clusters getting younger towards the centre. 
The 500 Myr-old clusters are believed to have formed during the first
encounter of the parent galaxies.  It is predicted that, following the
first encounter, the two galaxies separated to merge later and form the
younger clusters observed today (Mihos \& Hernquist 1994).  Comparison
with the 6 cm radio continuum maps of Neff \& Ulvestad (2000) indicates
that the optical detection of very young star
clusters in the Antennae is incomplete at
the 15 per cent level (Whitmore \& Zhang 2002). 

This age distribution has also been observed for star clusters in other
starburst galaxies, either when looking at distinct 
areas of a given galaxy or all
across a specific galaxy.  For example, in M82 (de Grijs et al.  2001)
and NGC 7252 (Miller et al. 1997) typical cluster ages 
decrease from 750 Myr to $<$ 10 Myr in going from
the outer regions inwards.  Significantly younger
clusters ($4 - 20$ Myr) have been resolved in NGC 1569 (Hunter et al. 
2000), NGC 3597 (Forbes \& Hau 2000), NGC 5253 (Schaerer et al.  1997)
and NGC 7673 (Homeier et al.  2002).  On the contrary, only clusters
older than 250 Myr were found in NGC 3921, the remnant 
of a merger that took place about 0.7~Gyr in the past 
(Schweizer et al.  1996). 

Young ($3 - 6$ Myr) and massive ($10^5 - 10^6$ M$_{\odot}$) star clusters
have been detected at the centre of more ``normal'', 
quiescent galaxies,
such as spirals with exponential (as opposed to $R^{1/4}$) bulges
(Carollo et al.  1997), late-type spirals (B\"oker et al.  2002) and
barred spiral galaxies (Colina et al.  2002, Gonzalez Delgado et al. 
2002), which are experiencing local and spatially confined 
starbursts.  Irrespective of the morphology
of their hosts, nuclear clusters in many cases dominate the radiation
field in the galactic centre, even when an AGN is present (Leitherer 2000).

Star formation is particularly enhanced in ULIGs, as they contain $10^8
- 10^{9.5}$ M$_{\odot}$ of molecular gas (Evans et al.  2001).  They are
typically the merger of two disk galaxies of nearly equal mass (mass
ratio $<$ 3:1, Sanders 2001).  HST/WFPC2 and NICMOS imaging has revealed
that $5 - 20$ percent of the known ULIGs are multiple mergers, probably
originating in compact groups of galaxies (Cui et al.  2001, Bushouse et
al.  2002).  Irrespective of their multiplicity, ULIGs generally show
spiral structure in their central regions, and circumnuclear bright
knots usually identified with concentrations of stars younger than $10^7
- 10^8$ yr (Scoville et al.  2000, Surace et al.  2000, Farrah et al. 
2001).  At optical wavelengths, ULIGs are characterised by nuclear
spectra which are either Seyfert-like (consistent with the presence of
an AGN or QSO), LINER-like or H{\sc ii}-region-like.  In these latter
cases, star clusters and/or H{\sc ii} regions seem to be the dominant
source of ionisation (Veilleux et al.  1999). 

The star clusters detected in ULIGs with ``cold'' and ``warm'' 
FIR colours reflecting the mean temperature of their 
radiating dust components (see Scoville et al. 2000) are the least
well-studied so far (except for the Antennae clusters) for a number
of reasons. First, they
suffer high extinction intrinsic to their parent galaxies. Secondly,
their host galaxies are found at larger distances than the best-studied
starburst galaxies and hence even the high angular resolution 
currently available with HST may not suffice. 
Nevertheless, star clusters in ULIGs are particularly important for
understanding star formation and its interplay with galaxy interactions,
since they reside in galactic systems at the early phases of their
merging process.  One way to shed light on the cluster properties in
ULIGs is to analyse the cluster properties in ``warm'' ULIGs where
stellar winds and supernova explosions have removed a significant
fraction of dust from the central region of the merger, making star
clusters accessible to optical--near-infrared (near-IR) observations. 

At a distance $D \simeq 98$ Mpc (for $H_0 = 75$ km s$^{-1}$ Mpc$^{-1}$),
NGC 6240 is one of the nearest and best-studied luminous infrared
galaxies (LIGs) with an infrared luminosity of $L_{\rm IR} = (3 - 11)
\times 10^{11}$ L$_{\odot}$ (Wright et al.  1984).  An infrared
luminosity greater than 10$^{12}$ L$_{\odot}$ is the criterion for
labelling a galaxy an ULIG (Sanders \& Mirabel 1996).  NGC 6240 is
therefore at the faint limit of the ULIGs class and can be
classified as a ``warm'' ULIG.  Its highly disturbed morphology is
suggestive of NGC 6240 being a merger of two massive disk galaxies
(Zwicky et al.  1961, Fosbury \& Wall 1979, Tacconi et al.  1999); the
colours of the nuclear and circumnuclear stellar populations indicate
that the merger has possibly been ongoing for the past $\simeq$ 1 Gyr
(Genzel et al.  1998, Tecza et al.  2000).  The gravitational
interactions as well as disturbances induced by star formation are
responsible for large-scale dust lanes, loops, shells and tails
extending out to $\sim 30$ kpc. 

The central region of the merger is
characterised by a double nucleus, with the two 
visible nuclei separated by
$1.''5 - 2''$ (0.7-1~kpc), on average, and their 
projected distance increasing at shorter
wavelengths (Bryant \& Scoville 1999, Scoville et al.  2000).  The H$_2$
and CO emission is found to peak in between the nuclei, possibly in a
thick disk structure, and slightly closer to the southern nucleus (Solomon 
et al. 1997, van
der Werf et al.  1993, Ohyama et al.  2000).  This offset may be due to
the interaction of the molecular gas with an expanding shell centred on
the southern nucleus (Rieke et al.  1985).  The double nucleus of NGC
6240 has been resolved into a number of compact star clusters or highly
excited H{\sc ii} regions with sizes of $0.''1 - 0.''2$, i.e.  50 to 100
pc, (cf.  Barbieri et al.  1993, Rafanelli et al.  1997).  These are the
dominant ionising sources in the nuclei (Lira et al.  2002)
and are also believed to drive a large-scale outflow (Heckman 
et al.  1987).  A nuclear
AGN component is also present, but obscured at optical-to-radio
wavelengths, and responsible for the galaxy's X-ray emission. 
Spectroscopy of the nuclei reveals an optical LINER-like spectrum with
significantly enhanced H$_2$ and [FeII] line emission in the near-IR,
very strong CO absorption bands and rather weak H recombination lines
which imply the presence of a Seyfert-like AGN (DePoy et al.  1986,
Heckman et al.  1987 and Carral et al.  1990). 

In this paper, we aim to estimate the age, mass and reddening of the
star clusters in order to discuss the star formation in NGC 6240 in
comparison with other well-studied merging systems.  In a sense, we want
to take ``advantage'' of NGC 6240 being a ``warm'', and thus 
relatively unobscured, ULIG to explore the
star formation modes in such gas-rich and dynamically extreme systems,
which are the ``embryos'' of the common spiral and elliptical galaxies. 
The ultimate question we want to address is whether the star formation
mechanisms in ULIGs are significantly different from what is observed in
normal galaxies. 

\section{Data processing}
\subsection{Observations}

We have made use of archive images taken with HST/WFPC2 as part of the
GO proposal 6430 (PI: R. van der Marel).  Their dataset rootname, filters
and exposure times are listed in Table 1. 

\begin{table}
\begin{minipage}{\hssize}
\caption{The log of NGC 6240 images taken with HST/WFPC2.}
\begin{tabular}{clr}
\hline
Dataset    & Filter          & Exp. Time (s)\\
\hline
u4ge010..  & F450W -- broad B        & 3  $\times$ 700.00 \\
           & F547M -- medium V       & 400.00, 800.00 \\
           & F656N -- H$\alpha$ cont.& 2 $\times$ 400.00 \\
           & F673N -- H$\alpha$      & 3 $\times$ 700.00 \\
           & F814W -- I              & 3 $\times$ 400.00 \\
\hline
\end{tabular}
\end{minipage}
\end{table}

We retrieved the WFPC2 images already pipeline-processed, i.e. 
corrected for bias, dark current and flat field.  We registered the
WFPC2 frames to the spatial grid of the images taken with the F450W
filter, by simply measuring the position of several stars ($\ge 5$) and
their relative shifts in (X;Y) among the available datasets.  The (X;Y)
shifts were then applied with the IRAF routine IMSHIFT.  Once aligned,
the images acquired with the same filter were combined with the STSDAS
task CRREJ to clean cosmic rays and sum individual exposures together
to improve their signal-to-noise (S/N) ratio. 

\subsection{WFPC2 photometry}  

Point-like bright sources could easily be identified by means of a
visual inspection given their small number.  Their positions were
measured with the IRAF routine IMEXAMINE.  These were used as the input
coordinates for PHOT in DAOPHOT.  Photometry of these compact sources
was performed assuming an aperture radius of 3 pixels in the PC images
and 2 pixels in the WF2 and WF4 exposures.  A median sky was computed
over an annulus of 5-to-8 pixels, thus chosen to be small and quite
close to the source because of the 
highly variable background of the galaxy. 

Aperture fluxes were corrected for the charge transfer (in)efficiency
(CTE) following the recipes given by Whitmore \& Heyer (1998).  Aperture
corrections to the standard 5 pixel aperture were determined from two
bright and isolated objects in the PC and WF chips, and subsequently
applied to the measured fluxes.  At this stage, fluxes were first
transformed into the WFPC2 synthetic magnitude system and subsequently
into the Johnson magnitude system using the zero points and the colour
equations listed in Holtzman et al.  (1995). 

The photometric errors are plotted in Figure 1 for the three broad-band
filters as a function of magnitude for the clusters detected on the PC
(filled dots) and WF (open dots) chips.  Flux uncertainties become
greater than 0.2 mag at an observed magnitude fainter than $\simeq$ 24.5
mag.  We also notice that the photometric errors derived for the
clusters in the tails (open dots in Figure 1) are larger than for the
clusters in the main body at fainter magnitudes.  This is probably due to 
different S/N ratios and background noise between the PC and
the WF chips.

\begin{figure}
\centerline{
\psfig{figure=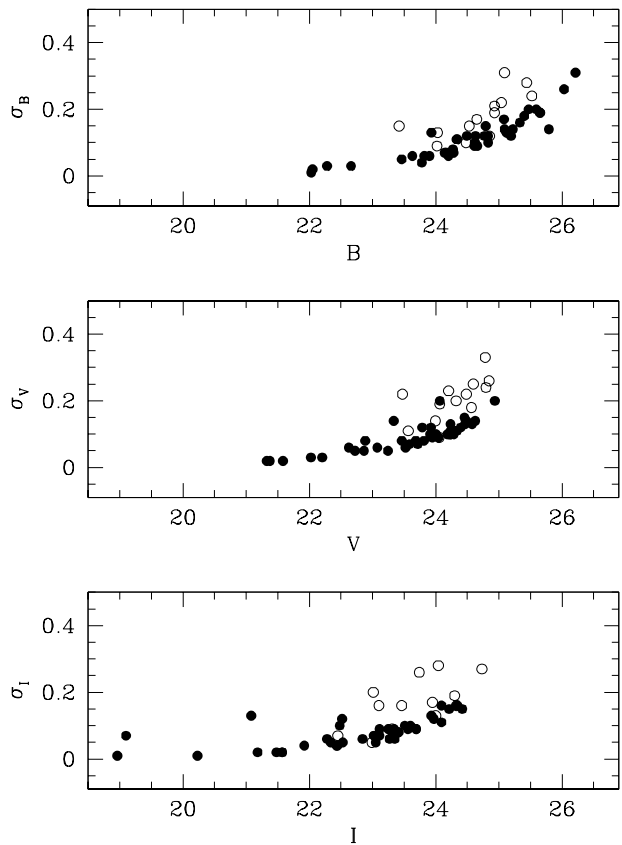,width=1.0\hssize}
}
\caption{Photometric errors as a function of magnitude for
the WFPC2 broad-band filters, F450W $(B)$, F547M $(V)$ and F814W $(I)$. Filled
dots are the clusters in the main body of NGC 6240, while open dots represent
the clusters detected in the galactic tails.}
\end{figure}

The limiting magnitudes of our cluster sample are 26.1, 25 and 24.5 in
$B, V$ and $I$ band, respectively.  They are associated with a S/N ratio
in the aperture flux of 3.2, 2.0 and 1.8 respectively, implying 
a completeness limit ($>$ 90$\%$) of V $\sim$ 24.5 (e.g., Figure 4).  Under the
assumption that the clusters can be assigned a B0 spectral type with
E$(B-V)$ = 0.5 on average, the WFPC2 Exposure Time Calculator computes
for the exposure times and the number of exposures for each filter
(``CR-SPLITs'') in Table 1 the following S/N values: 2.2 at $B = 27$,
2.7 at $V = 26$ and 2.4 at $I = 25.5$. These represent the 
most optimistic performance of WFPC2 without the complex 
backgrounds of NGC~6240, and confirms that the above 
estimate of sample completeness is reasonable 
for the exposure times of the observations. However, as the observations 
are relatively shallow and the distance of NGC 6240 is 98 Mpc, we 
see that we cannot sample very far 
down the star cluster luminosity function.
 
At the NGC 6240 redshift of $z$ = 0.0245 the H$\alpha$ line falls within
the spectral range of the F673N filter.  The adjacent F658N filter
samples the continuum blueward of the H$\alpha$ emission.  Therefore, we
subtracted the F658N image from the F673N to get the net H$\alpha$
emission from NGC 6240.  Both images were scaled by their exposure times
and filter bandwidths prior to subtraction. 

\section{The galaxy morphology}

The colour mosaic of NGC 6240 is shown in Figure 2, where blue is assigned to the
F450W image, green to the F547M and red to the F814W exposure. The H$\alpha$
continuum-subtracted image appears here in yellow. 

\begin{figure}
\caption{Colour composite of NGC 6240. The blue colour represents the F450W image,
the F547M image is shown in green and red is assigned to the image in
F814W. The H$\alpha$ emission is shown in yellow. North is up and East to the left.} 
\end{figure}

The broad-band filters will be contaminated by nebular emission.  We
have redshifted the template spectrum of starburst $\#$ 1 [E$(B-V) <
0.1$ mag, which shows the most prominent [OIII] 5007 emission among the starburst templates in Kinney 
et al.'s (1996) atlas] to $z$ = 0.0245 and
overplotted it on the sensitivity functions of the F450W, F547M and
F814W filters in Figure 3 to illustrate the influence of emission lines. 

\begin{figure}
\centerline{
\psfig{figure=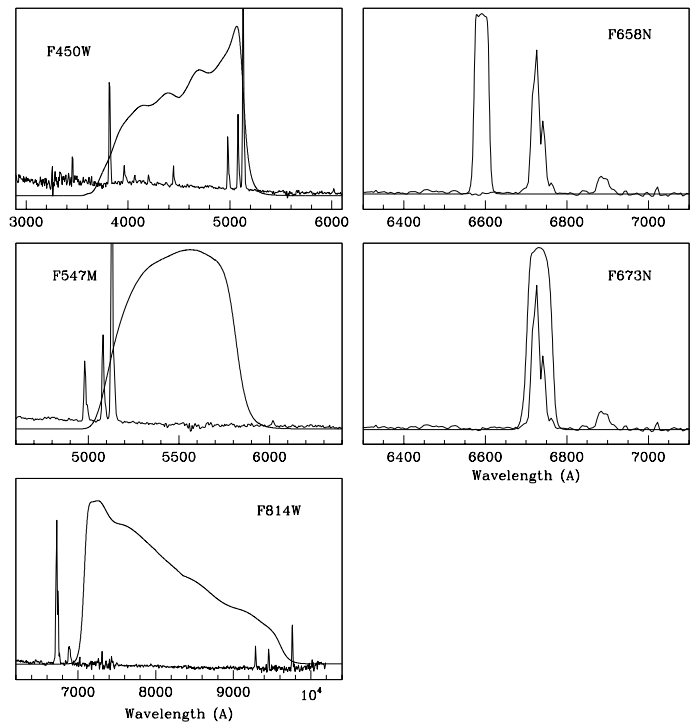,width=1.1\hssize}
}
\caption{Template spectrum of starburst $\#$ 1 [E$(B-V) < 0.1$ mag] in Kinney
et al.'s (1996) atlas, overplotted on the
WFPC2 filters used to observe NGC 6240. The spectrum is redshifted by the
same redshift measured for NGC 6240, i.e. {\it z} = 0.0245. Compared to 
this template, NGC~6240 has weak [OIII] emission (Kennicutt 1992).}
\end{figure}

It turns out that nebular emission lines mostly contaminate the F450W
filter, with the [OIII] doublet at $\lambda\lambda$ = 4959, 5007 \AA\/
falling at the peak of the filter response curve.  The same doublet is
intersected by the F547M filter where the filter response is very low,
so that the [OIII] doublet flux in the F547M filter is a factor $\sim$ 3
smaller than in F450W.  Fortunately, the integrated spectrum 
found by Kennicutt (1992) indicates the average [OIII] emission 
line equivalent width is much smaller than in the starburst template 
spectrum. However, there may be small regions where this line is strong 
enough to influence the F450W flux. 
The F814W filter is essentially free of nebular
emission. Note that our assumptions generally are conservative in that 
much of NGC~6240 has weaker line emission than the template starburst 
spectrum in Figue 3.

In all three broad-band images, the central area of NGC 6240 (contained
in the PC frame) is dominated by a dust lane $\simeq$ 25$''$ (12 kpc at
$D = 98$ Mpc) in length and $\simeq$ 2 kpc in width, and with its
major axis at PA $\simeq$ 24$^\circ$.  The centre of the dust lane is at
about 4$''$ (1.9 kpc) NW from the centre of the galactic double nucleus. 
The dust lane is embedded in gas which mimics two lobes aligned with
the major axis of the dust lane.
A gaseous arm departs from the double nucleus of NGC 6240 at PA
nearly 90$^\circ$. 

Two galactic tails (detected in the WF2 and WF4 chips) develop East of
the dust lane and merge with the double nucleus.  The northern tail is
more extended than the southern: their length is 58$''$ (28 kpc) and
40$''$ (19 kpc), respectively.  The northern tail shows dust patches
which are absent in the southern tail.  According to the morphological
scheme of Surace et al.  (1998, 2000) and Surace \& Sanders (1999), NGC
6240 can be included in their class III, i.e.  a pre-merger system with
tidal tails and a double nucleus at a separation of less than 10 kpc. 

The image in the H$\alpha$ line has a different
morphology.  The major axis of the H$\alpha$ emission is perpendicular
to the dust lane and along it a superbubble develops towards the West. 
This is the main feature of NGC 6240 in the light of H$\alpha$.  The
H$\alpha$ emission is concentrated on the northern and southern rims of
this superbubble; they extend for about 9$''$ (4.3 kpc) but do not merge
into each other.  The superbubble perimeter is thus incomplete.  A
significant amount of radio emission at 20 cm is associated with this
western superbubble (Colbert et al.  1994).  The radio spectrum is
characterised by a steep slope ($\alpha \simeq$ 1.0) which does not arise 
from relativistic electron inverse Compton losses on starlight, 
and therefore may be due to expansion of the relativistic plasma in a 
galactic wind.
A smaller bubble, 5$''$ (2.5 kpc) in diameter,
nests with the northern rim of the superbubble.  A second superbubble is
visible South of the double nucleus; two rims are detected extending out
6$''$ (2.8 kpc) but do not complete the perimeter of the superbubble. 
Finally, an H$\alpha$ filament is aligned along the gaseous arm at PA
$\simeq$ 90$^\circ$ and is about 6$''$ (2.8 kpc) long.  The small-scale
structure of the H$\alpha$ emission is characterised by condensations
and bright knots. 

Imaging and spectroscopy of the H$\alpha$ filaments in NGC 6240 were
obtained by Heckman et al.  (1987) who noticed their similarity to those
in M 82 and suggested that they are the result of a galactic wind
ionising gaseous tidal debris surrounding the galaxy.  These authors
proposed a model in which the kinetic energy released by supernovae in
the starburst region provokes high-speed gas flows which shock the local
interstellar medium and form a central cavity of hot gas.  The hot gas
flows then outward and along the rotation axis of the galaxy, sweeping
and shocking the gas clouds from the merging galaxies.  Heckman et al. 
(1990) demonstrated that this model reproduces the observed size of the
filaments, and hence ruled out the possibility that an AGN drives the
galactic wind in NGC 6240.

In summary, NGC~6240 resembles M82 in terms of its near edge-on orientation, 
physical scale of the starburst region, and presence of large outflows. 
However, its order of magnitude more powerful starburst was triggered by 
a merger, while that in M82 appears to be associated with a relatively 
mild interaction. Furthermore, M82 appears to lack an AGN, but the 
ongoing merger in NGC~6240 has stimulated modest nuclear activity 
(see Beswick et al. 2001; Lira et al. 2002 and references therein). 
M82 and NGC~6240 therefore provide good opportunities to investigate how 
the magnitude and triggering of starbursts influence the resulting 
population of SSCs.

\begin{figure}
\centerline{
\psfig{figure=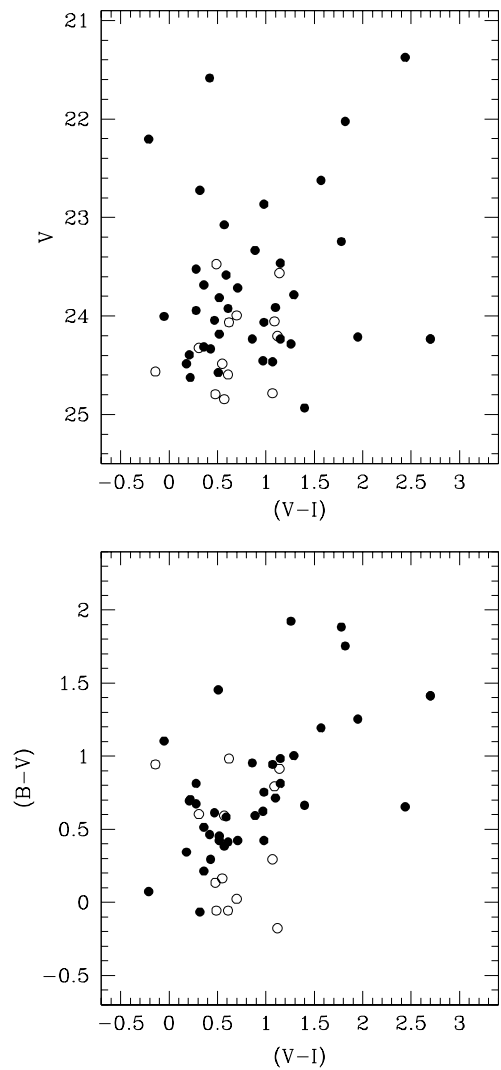,width=0.8\hssize}
}
\caption{Two-colour diagram for the clusters in the main body of
NGC 6240 (filled dots) and in the galactic tails (open dots).}
\end{figure}

\section{The star cluster population}

A total of 41 star clusters were detected on the PC images, except in the
central $1'' \times 2''$ area where the two galactic cores lie.  Their
average FWHM is 2.4 pixels; given that the stellar PSF measured in the
Galactic globular cluster NGC 7078 (WFPC2 parallel programme ID 8805, PI
Casertano) in the same filters has FWHM = 1.9 pixels, we conclude that
the clusters on the PC images are very marginally resolved.  In addition, 13
clusters were identified in the WF2 and WF4 images.  Their FWHM is 1.7
pixels on average, which is comparable to the stellar PSF FWHM of 1.3
pixels measured in NGC 7078, and hence the clusters in the WF images are
not resolved. A table collecting the coordinates and the photometry of the 
clusters in the main body and tails of NGC~6240 is available at
http://www.ast.cam.ac.uk/STELLARPOPS/Starbursts/. 
The photometric accuracy of the whole cluster sample is less
than 0.2 mag; WFPC2 stellar population studies suggest
that incompleteness at $\Delta$m = 0.2 becomes significant in non-crowded fields.

We have plotted both cluster samples in Figure 4, in the $(V,I)$
colour-magnitude diagram and in the $(B,V,I)$ two-colour diagram. 
Magnitudes and colours have been corrected for the Galactic reddening
towards NGC 6240 of E$(B-V) = 0.076$.  The extinction law by Savage \&
Mathis (1979) was adopted.  Clusters from the PC images are represented by filled
dots, while clusters from the WF2 and WF4 chips are shown as open dots. 
On average, the WF clusters appear to be fainter in $V$ and bluer in
$(B-V)$ and $(V-I)$.  This is probably due to a lower reddening in the
tails compared to the main body of the galaxy. 

\begin{figure}
\centerline{
\psfig{figure=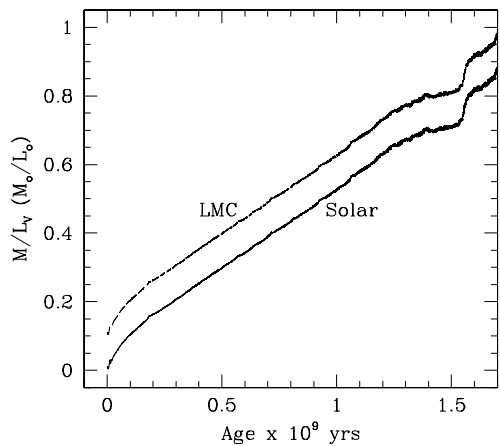,width=0.8\hssize}
}
\caption{The cluster M/L ratio as a function of age, computed from two
STARBURST99 models at solar and LMC-like metallicity respectively,
and with a cluster mass of 10$^5$ M$_{\odot}$. The distribution
obtained for a LMC-like metallicity has been here shifted by 0.1
in M/L, but, in reality, overlaps perfectly with the distribution
computed for a solar metallicity.}
\end{figure}

For both cluster samples, observed magnitudes and colours were compared
to STARBURST99 simple stellar population 
models (Leitherer et al.  1999) to estimate cluster
masses and ages.  In particular, since no metallicity has been measured
yet for NGC 6240 (McCall et al.  1985 predicted a metallicity about
20 per cent solar), the STARBURST99 models were computed for both Z =
0.20 Z$_{\odot}$ and Z = Z$_{\odot}$.  The models assume a single
episode of star formation, characterised by a Salpeter IMF from 0.1
M$_{\odot}$ to 100 M$_{\odot}$ and standard mass loss. 
The M/L ratio has been derived as a function of age from both models,
assuming a cluster total mass of 10$^5$ M$_{\odot}$. These ratios
are plotted in Figure 5, with the ratio computed for a LMC metallicity
shifted by $\+$0.1 from the solar M/L values. In reality the two
distributions overlap, indicating that  
the two metallicities do not produce detectable differences
in the clusters M/L for a given age. Anders \& Fritze-von Alvensleben (2003)
have shown that the inclusion of nebular emission in population
synthesis models introduces a dependence of M/L on the cluster metallicity
at the youngest ages. In the case of NGC 6240 whose average [OIII] line
equivalent width is much smaller than in the starburst template spectrum
of Figure 3, we expect the effect of nebular emission not to be significant. 

Unfortunately, the H$\alpha$ image is not deep enough to measure the
integrated H$\alpha$ flux for the clusters in the NGC 6240 main body and
use it together with the observed colours to constrain cluster ages, as
Whitmore \& Zhang (2002) have done for the clusters in the Antennae. 
This method can be applied when clusters are detected in the same
range of age as those in the Antennae, which appears to be the case 
for many of the star clusters in NGC 6240, as we discuss in Sect. 5.

The luminosity function for all clusters
in our sample is plotted in Figure 6 in terms
of the observed V$_0$ magnitude corrected for Galactic extinction
towards NGC 6240. Absolute magnitudes M$_{V}$ have been computed
by scaling V$_0$ by the galaxy distance modulus (34.96 mag).
No attempt has been made to correct for internal interstellar
obscuration, which is poorly determined for individual star clusters.
The luminosities in Figure 6 thus represent lower bounds.

\subsection{The clusters in the main body of NGC 6240}

The STARBURST99 models for Z = 0.20 Z$_{\odot}$ are plotted on the
$(V,I)$ colour-magnitude diagram for different total cluster masses in
Figure 7. Here, each evolutionary track (solid line) is represented by
its more significant mesh points: the maximum $V$ luminosity (reached at
an age of 4 $\times$ 10$^6$ yr), the maximum colour excursion (occurring
between 9.5 $\times$ 10$^6$ yr and 1.8 $\times$ 10$^7$ yr) and a fading
phase at an age of about 1.3 $\times$ 10$^8$ yr.  Each point is reddened
by steps of 0.2 mag in E$(B-V)$, the reddening increasing from 
0.  to 0.6 mag for cluster masses $\leq$ 10$^5$ M$_{\odot}$ and from 0.2
to 1.0 for a cluster mass of 10$^6$ M$_{\odot}$.  The extinction 
law here adopted to treat the cluster intrinsic reddening is Calzetti's (2001). 

\begin{figure}
\centerline{
\psfig{figure=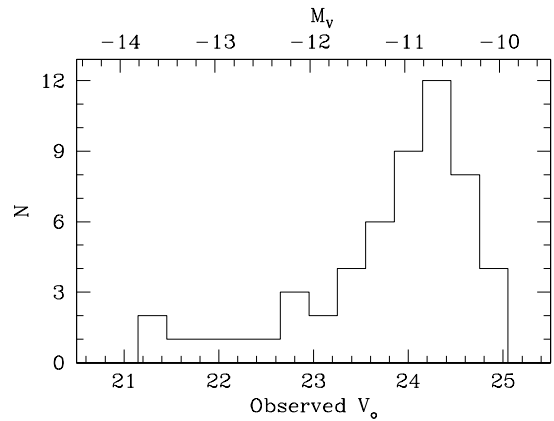,width=0.45\hdsize}
}
\caption{The luminosity function in V magnitude (corrected for
the galactic estinction towards NGC 6240) as derived for
the main body $\+$ tail sample of clusters.}
\end{figure}

Magnitudes and colours for models 
at the same age but different reddening values
are connected by dashed lines.  The track
computed for $M_{\rm cluster}$ = 10$^4$ M$_{\odot}$ fits the observed
distribution of clusters poorly for E$(B-V) <$0.2. 
The bulk of the clusters in the main body of NGC 6240 are equally well
reproduced by the tracks at $M_{\rm cluster}$ = 5 $\times$ 10$^4$ and
10$^5$ M$_{\odot}$.  A small number of clusters redder than $(B-V)
\simeq$ 1.5 mag, is matched only by the STARBURST99 model calculated for
$M_{\rm cluster} \ge$ 10$^6$ M$_{\odot}$.  Clearly, on a finer grid of
evolutionary tracks based on smaller E$(B-V)$ steps, clusters can easily
shift in age between the $4 - 7$ and $9 - 18$ Myr ranges, but clusters
older than 1.8 $\times$ 10$^7$ yr can be readily distinguished. 

Although the F450W filter is contaminated by possible nebular emission,
we present the $(B-V)$ colour-magnitude diagram in Figure 8, which is
based on a set of four STARBURST99 models of different total mass (from
10$^4$ to 10$^6$ M$_{\odot}$) and for Z = 0.20 Z$_{\odot}$.  The symbols
are as in Figure 7.  The best agreement is found for the mass interval $(0.5 -
1) \times$ 10$^5$ M$_{\odot}$.  

\begin{figure*}
\centerline{
\psfig{figure=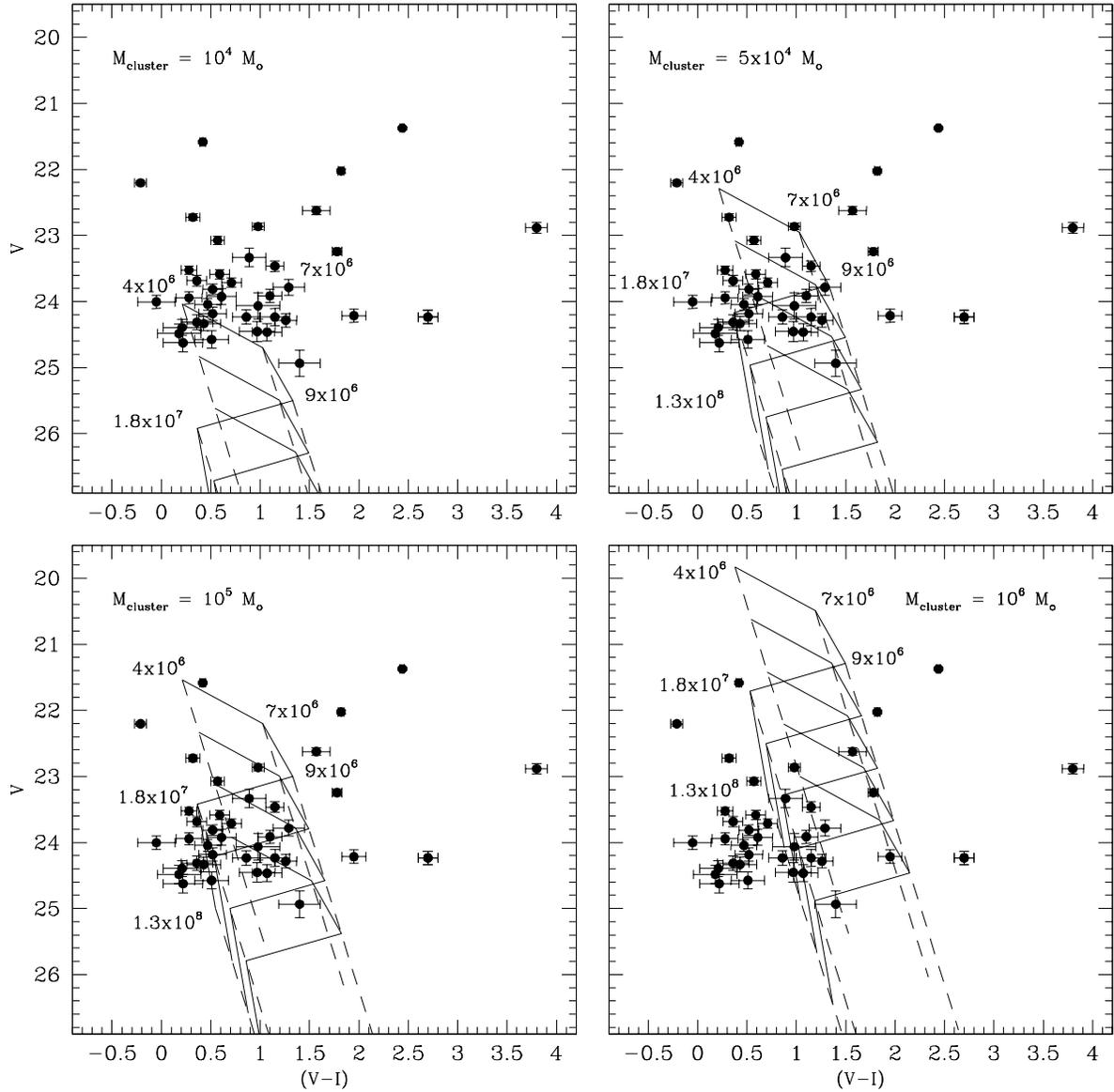,width=0.9\hdsize}
}
\caption{$V$ vs.  $(V-I)$ colour-magnitude diagram of the clusters in
the main
body of NGC 6240, compared to the evolutionary tracks obtained for
cluster masses of 10$^4$ M$_{\odot}$, 5 $\times$ 10$^4$ M$_{\odot}$,
10$^5$ M$_{\odot}$ and 10$^6$ M$_{\odot}$ and reddened in steps of
E$(B-V)$ = 0.2 mag.}
\end{figure*}

\begin{figure*}
\centerline{
\psfig{figure=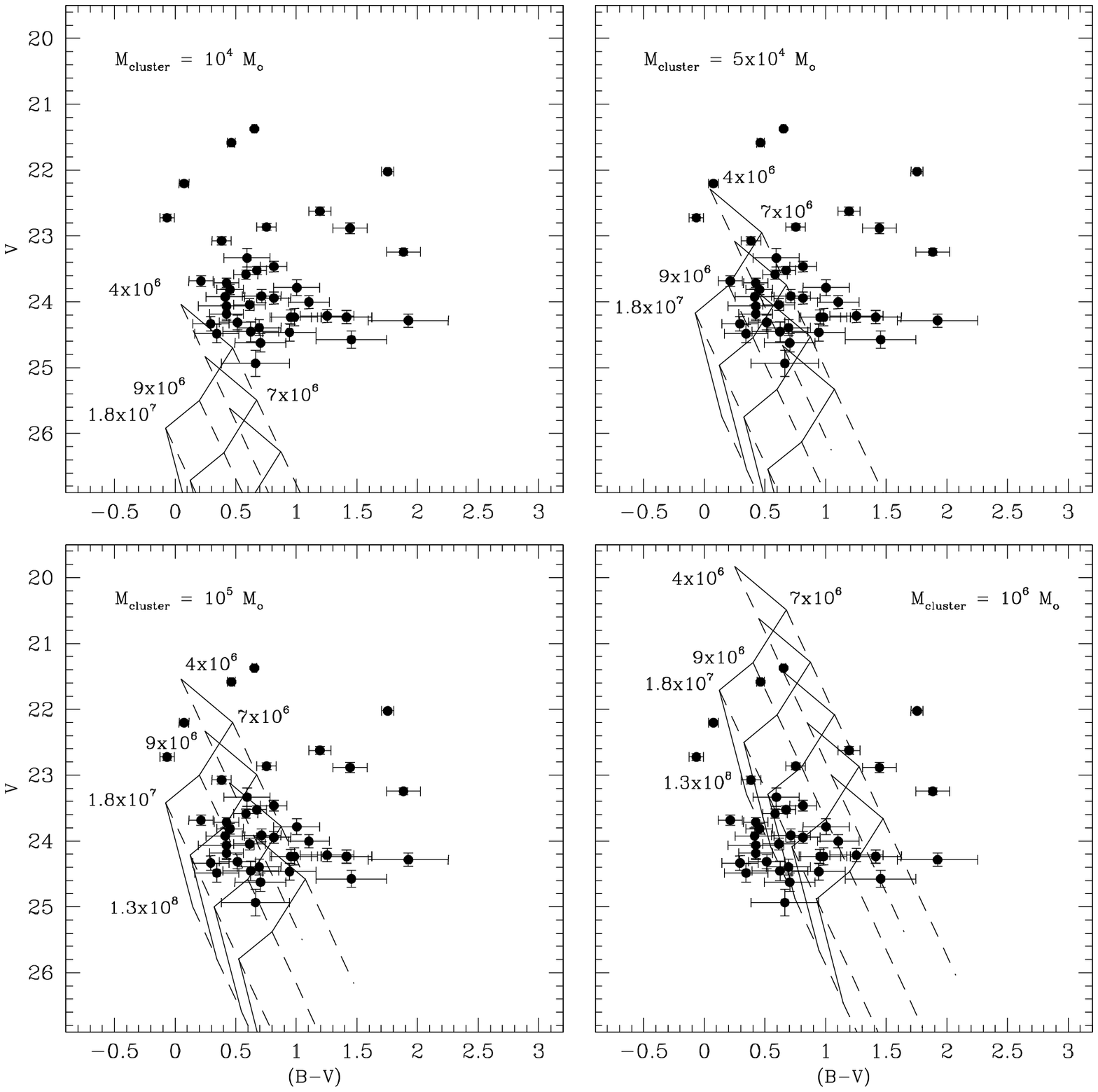,width=0.9\hdsize}
}
\caption{$V$ vs.  $(B-V)$ colour-magnitude diagram of the clusters in the main
body of NGC 6240, compared to evolutionary tracks obtained for
cluster masses of 10$^4$ M$_{\odot}$, 5 $\times$ 10$^4$ M$_{\odot}$,
10$^5$ M$_{\odot}$ and 10$^6$ M$_{\odot}$ and reddened in steps of
E$(B-V)$ = 0.2 mag.}
\end{figure*}

The assumption of a larger cluster mass, $M_{cluster} = $10$^5$
M$_{\odot}$, slightly improves the agreement of observed distribution of clusters 
with the evolutionary models. The clusters shift towards 
models with an older
age of $\sim$ 1.8$\times$ 10$^7$ yr and a higher mean reddening value of
E$(B-V) \sim$ 0.6 mag.  Clusters redder than $(B-V) \simeq$ 1.2 mag can only
be matched by models with a mass greater than 10$^6$ M$_{\odot}$.  These appear
to be as young as 4 to 7 Myr and span a rather large interval of E$(B-V)$
values, between 0.2 and 1.2 mag.  As Figure 5 shows, the above results
also hold when a solar metallicity is assumed for NGC 6240.

The luminosity function is plotted in Figure 6 in terms
of the observed V$_0$ magnitude corrected for the galactic extinction
towards NGC 6240. Absolute magnitudes M$_{V}$ have been computed
by scaling V$_0$ by the galaxy distance modulus (34.96 mag), but
no attempt has been made to correct for intrinsic reddening which has
been poorly estimated.

\subsection{Model Fitting: the clusters in the NGC 6240 main body}

In an attempt to better quantify ages and reddening values of the clusters
in the main body of NGC 6240, one of us (AP) developed 
a program to fit the observed $(B-V)$ and
$(V-I)$ colours, dereddened by the Galactic E$(B-V)$ = 0.076, to an
evolutionary track computed for a total cluster mass of 10$^5$
M$_{\odot}$ and an LMC metallicity.  Except for stochastic 
effects associated with small numbers of very luminous stars, model 
cluster colours are not mass sensitive, but mainly
depend on age and reddening.  Since the cluster luminosity scales
linearly with the logarithm of its mass, we have computed the difference
between the observed $V$ magnitude and the model apparent $V$ given for
a specific age and reddening and used this to derive the actual cluster
mass via the equation: 10$^{-0.4(V_{\rm obs}-V_{\rm th})}$ $\times$
10$^5$ M$_{\odot}$.  

Specifically, for a given E$(B-V)$ the $(B-V)$ and $(V-I)$ colours
computed for each time point of the evolutionary track have been
reddened and a $\chi^2$ has been calculated as the sum of the
differences between observed and theoretical/reddened colours normalised
by the observational errors:
\par\noindent
$\chi^2$ = $\left({(B-V)_{\rm obs} - (B-V)_{\rm th} \over
\sigma_{BV}}\right)^2$ $+ \left({(V-I)_{\rm obs} - (V-I)_{\rm th}\over
\sigma_{VI}}\right)^2$. 
\par\noindent
The assumed E$(B-V)$ was varied from 0.  to 5.  mag with increasing
steps of 0.01 mag and it was applied following Calzetti's (2001)
reddening law.  The final result of this procedure is, for each cluster,
a grid of $\chi^2$ values depending both on reddening and age, which are
associated with a cluster mass as derived above.  From each cluster
grid, only the ages, reddening values and cluster masses corresponding
to $\chi^2 \leq$ 1 were kept; the fit to the evolutionary
track involves two observed parameters [$(B-V)$ and $(V-I)$] and two
unknowns (age and reddening).  {\it In other words, each cluster is
characterised by a range in  age, reddening and mass, 
that are consistent with the data.} 
\par\noindent

We have combined the cluster ranges spanning all of the age, reddening
and mass parameters in a single histogram.  The number of clusters
associated with each bin was normalised by the total number of clusters
detected in main body of NGC 6240.  The age, reddening and mass
statistical distribution
histograms are plotted in Figure 9 with a solid line for the clusters in
the galaxy main body; they represent the probability that any cluster
has a given age, reddening or mass. 

The upper panel of Figure 9 gives the age probability function of the
cluster sample; the distribution covers an age range between 10$^6$ and
10$^{10}$ yr with 60 per cent of the clusters aged between 2 $\times$
10$^6$ and 10$^9$ yr.  Only 50 per cent of the clusters could be either
younger than 2 Myr or older than 1 Gyr.  A peak of 70 per cent appears
in the distribution at an age of about $5 - 6$ Myr which could partially
be due to luminosity selection effects, as younger clusters are brighter
and thus more easily detected. In order to assess how luminosity selection 
effects affect the probability functions in Figure 9, we have traced, from
the STARBURST99 models, the cluster brightness as a function of age per 
given mass up to the epoch 
when the cluster fades below the limiting magnitudes of our sample. The
results are shown in Figure 10, where the cluster mass is plotted as
a function of age: the dotted area indicates the mass and age of those
clusters which escape detection since they are fainter than the limiting
magnitudes of our sample.  The distribution in Figure 10 has
been computed for $E(B-V)$ = 0.4 mag, which is the peak of the reddening
distribution of the clusters in the galaxy main body (cf. bottom panel of Figure 
9). An higher reddening
reduces the cluster detection in NGC 6240 to
objects more massive and younger. The peaks of the age and mass distributions
for both the cluster samples in the galaxy main body and tails are plotted
in Figure 10, as filled and open dots respectively. Clusters at any age
with a mass smaller than $\sim$ 4 $\times$ 10$^4$ M$_{\odot}$ are not detected; 
clusters of increasing mass are detected at gradually older ages.
Therefore, the sharp decline of the mass distributions at M $\leq$ 4 $\times$ 
10$^4$ M$_{\odot}$ in Figure 9 is due to luminosity selection effects.

Clearly, the individual cluster ages in Figure 9 are poorly determined. The reason for this
becomes evident in Figure 11, where we have plotted the evolutionary
track used in our fitting procedure and superimposed the distribution of
the observed cluster colours, corrected for Galactic foreground
reddening.  The evolutionary track has been reddened by E$(B-V)$ = 0.0,
1.0 and 1.5 mag respectively (by applying Calzetti's reddening law) and
sample points at the same age have been connected with a thin solid
line. 
Only the clusters in the range $(B-V) \leq$ 1.0 and $(V-I) \leq$
0.8 mag  are fitted unambiguously by young ages ($\simeq$ 3 Myr)
at any reddening.  Redder clusters  intercept the evolutionary
tracks at a number of points, each at a different age (cf.  Whitmore \&
Zhang 2002). 

The middle panel of Figure 9 shows the mass probability function for the
clusters in the main body of NGC 6240 as the solid line.  It can be seen
that the cluster mass probability increases from 10$^4$ to
10$^5$ M$_{\odot}$ (since higher masses are less affected by luminosity
selection effects) where the mass distribution peaks with 70 per cent of
the clusters, and declines almost linearly in log$(M)$ down to a
probability of about 5 per cent of the clusters having a mass of
$\simeq$ 3 $\times$ 10$^8$ M$_{\odot}$. 
Figure 6 shows that the observed luminosity function is
approximately a power law. Figures 5, 7, and 8 indicate how for
a fixed cluster mass (in this case 10$^5$ M$_{\odot}$)
a cluster fades as it ages. Therefore, if differential
reddening were negligible and the clusters were coeval with
the same IMF, their mass distribution would follow a
power law as well. Unfortunately, the lack of extended multiwavelength
observations of NGC 6240 does not allow us to pin down accurate
values for the clusters age, mass and reddening, but it only
makes possible to draw a range of the most probable values
that cluster age, mass and reddening can have. When these ranges
are merged together to construct the probability function of
any of the above parameters (cf. Figure 9), their width (i.e. error bar)
has the effect of smearing the relationship between apparent luminosity
and cluster mass, and the probability function shows a possible
peak.

Similarly, Fritze-von Alvensleben (1999) has demonstrated
that evolutionary effects tend to broaden power-law luminosity
functions. Two-colour photometry is certainly not sufficient to
derive the {\it true} cluster mass distribution for a given
cluster age, only a mass range can be computed which is most
consistent with  the data. This is  a statistical rather than
a qualitative (usually adopted in other works) approach for the
comparison between data and evolutionary predictions.
The bottom panel of
Figure 9 shows the reddening probability distribution of the main body
clusters (solid line).  The majority of the cluster sample (80 per cent)
seems to be associated with E$(B-V) \simeq$ 0.5 mag and the remaining
clusters populate bins of higher reddening (up to 2 mag) at rapidly
decreasing probability. We expect this trend in an optically selected 
sample where highly obscured clusters will be difficult to observe.

\subsection{The clusters in the NGC 6240 tails}

The clusters detected in the tails of NGC 6240 (on both the WF2 and WF4
chips) are shown in the $(B,V,I)$ colour-magnitude diagram of Figure 12,
where evolutionary tracks (solid line) are plotted for two different cluster
masses following the nomenclature of Figure 7.  Similarly to Figure 7,
each model is reddened following Calzetti's (2001) extinction law, with
an E$(B-V)$ increasing from 0.  to 1.0 mag in 0.2 mag steps.  Mesh
points at the same age but different reddening are connected by dashed
lines.  The observed clusters are simply corrected for the Galactic
extinction measured towards NGC 6240. 

When compared to a model of 10$^4$ M$_{\odot}$, 50 per cent of the
clusters appear to have young ages, between 4 and 7 Myr and low
reddening, between 0.  and 0.2 mag.  The best fits to the observed
distribution of clusters are given by the evolutionary tracks computed
for cluster masses of about 1 $\times$ 10$^5$ M$_{\odot}$. 
The model for $M_{\rm cluster}$ = 10$^5$ M$_{\odot}$ reduces the
fraction of clusters with ages younger than 9 $\times$ 10$^6$ yr and
gives, in general, older ages, between 9 and 130 Myr, associated with
0.2 $\leq$ E$(B-V) \leq$ 0.4 mag.  

We computed age, mass and reddening probability functions for the
clusters in the tails in the same way as for the clusters in NGC 6240
main body.  These distributions are plotted in Figure 9 
as dashed lines. 
The age distribution in the top panel of Figure 9 spans a large
interval, from 10$^6$ to 10$^{10}$ yr, with a maximum probability
($\simeq$ 85 per cent) at an age of about 13 Myr.  The probability of
cluster ages older than this slowly decreases down to 25 per cent
at an age of about 1 Gyr. The clusters in NGC 6240 tails range in mass
between 7 $\times$ 10$^3$ and 10$^8$ M$_{\odot}$ (middle panel of Figure
9).  A mass of about 1.8 $\times$ 10$^5$ M$_{\odot}$ appears to be most
probable at the 90 per cent level. The reddening distribution (in the
bottom panel of Figure 9) peaks towards low E$(B-V)$ values, $\simeq$
0.2 mag, although a tail of decreasing probability extends up to
E$(B-V)$ = 2.0 mag (at the level of about 15 per cent). The combination
of a lower reddening with the limiting magnitudes of the cluster
sample  is likely to reduce the luminosity selection effects on
the mass distribution of the clusters in the tails, so that the
ascending part of their mass probability function between
10$^4$ and 10$^5$ M$_{\odot}$ appears more populated than for the more
reddened clusters in the galaxy main body.

\section{Discussion}

\begin{figure}
\psfig{figure=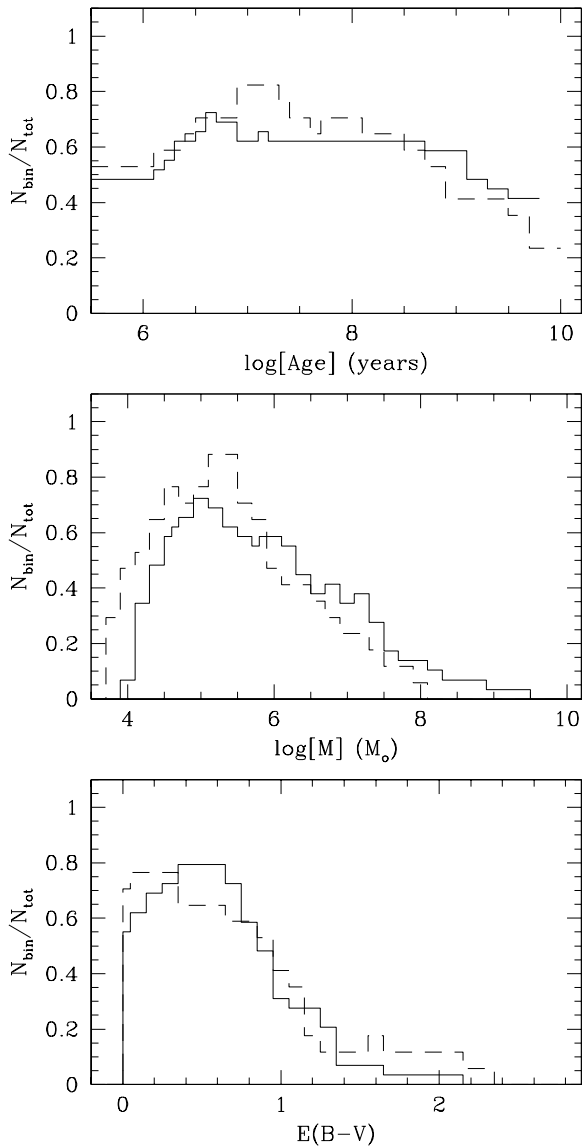,width=0.45\hdsize}
\caption{Probability distributions of age and mass in logarithmic units,
and reddening values on a linear scale.  The solid line represents the
clusters detected in the NGC 6240 main body, the dashed line the
clusters identified in the galaxy tails.}
\end{figure} 

\subsection{Star cluster masses and luminosities} 
A qualitative fit to the the observed visual 
luminosity distribution in Figure 
6 gives a power-law luminosity function N(L) $\sim$ L$^{-1.6}$ for -12 
$\leq$ M$_{V} \leq$ -10. This shallow function, biased towards bright
clusters, is likely to be a selection effect, as the relatively short 
exposures reduce the detection 
probability for clusters fainter than V$_0 \simeq$ 24.5. In addition,
the slope of a lower-law luminosity function has been shown by
Fritze-von Alvensleben (1999) to be likely affected by the age range
among the cluster sample.
On the basis of the data collected so far, the 
power-law index from fits to luminosity functions of 
SSCs associated with starburst galaxies usually lies in the
range between -1.7 and -2.1 (Whitmore 2000).
Some evidence of a steeper power-law index is suggested by 
the brighter clusters in NGC 3256 (Zepf et al. 1999), while 
the extreme value of -2.6 is derived for the Antennae's star clusters, whose
absolute V magnitudes span -12.9 and -10.4 (Whitmore et al.
1999).

\begin{figure}
\psfig{figure=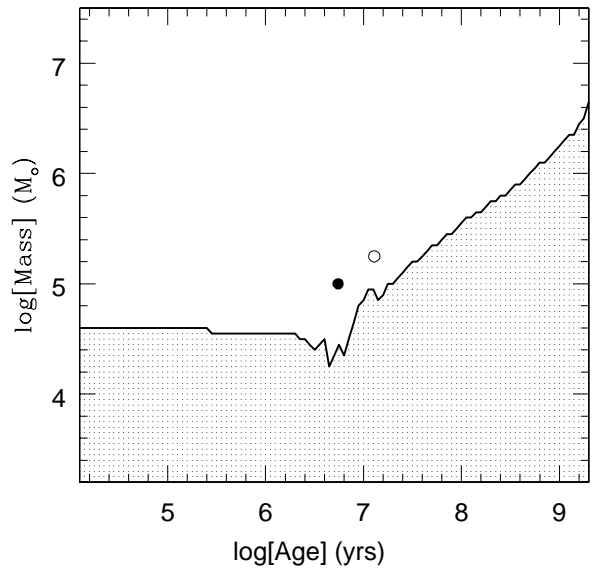,width=0.45\hdsize}
\caption{The mass and age ranges spanned by the clusters in our sample.
The dotted area gives the mass and age of clusters which escape detection
because they are fainter than
the limiting magnitudes of our sample. This distribution has been derived
for $E(B-V)$ = 0.4 mag, which is the peak of the reddening distribution
of the clusters in the NGC 6240 main body.}
\end{figure}

In the case of the Antennae, for which 
extensive multi-wavelength photometry is
available, Zhang \& Fall (1999) show that the star cluster mass function
is also roughly a power-law.  Unfortunately, such a wide photometric
coverage is not available for the majority of the young clusters observed 
in other galaxies, including NGC 6240, and therefore the form of their mass 
distribution can not be determined. This parameter plays a key role
in our understanding of the formation and destruction mechanisms of
young clusters as a function of their environment, and their possible
evolution into globular clusters. We therefore consider what we 
might learn from mean properties of the NGC~6240 star cluster system.

Figure 9 shows that ages between 2 $\times$ 10$^6$ and 10$^9$ yr
are applicable to more than 60 per cent of the detected star clusters (galactic
main body and galactic tails) with a hint that some clusters in the main
body of NGC 6240 could be younger (70 per cent at $5 - 6$ Myr), and the
clusters in the galactic tails slightly older (85 per cent at 13 Myr). 
The mass distributions indicate that more than 60 per cent of the observed 
cluster sample has a mass between 3 $\times$ 10$^4$ and 10$^6$
M$_{\odot}$ and a very small fraction of clusters (about 5 per cent)
could be as massive as few $\times 10^8$ M$_{\odot}$.  The most probable
cluster mass ($\ge$ 70 per cent) is $(1 - 2) \times$ 10$^5$ M$_{\odot}$. 
At the same cut-off level of 60 per cent, the cluster sample appears to
be reddened by E$(B-V)$ between 0.  and 0.8 mag; a reddening greater than
1.3, and up to 2 mag, is associated with 5 per cent of the clusters, 
as illustrated in Figure 11.

A finer way to look for radial gradients in age, mass and reddening is
to subdivide the cluster sample in the NGC 6240 main body in two annuli
centred on the nuclear source A1 identified by Barbieri et al.  (1993).
The inner annulus extends from R$_{\rm i}$ = 2$''$ to R$_{\rm o} <$
8$''$ (1 to 4 kpc) while the outer annulus covers the radial distance
range 8$'' \le$ R $\le$ 14$''$ (between 4 and 7 kpc).  We determined the
probability distributions for both annuli and compared them to those of
the cluster sample in the NGC 6240 tails.  The distributions are plotted
in Figure 13, where the inner annulus is the solid line, the outer
annulus the dashed line, and the clusters in the tails are represented
by a dot-dashed line. 

\begin{figure}
\centerline{
\psfig{figure=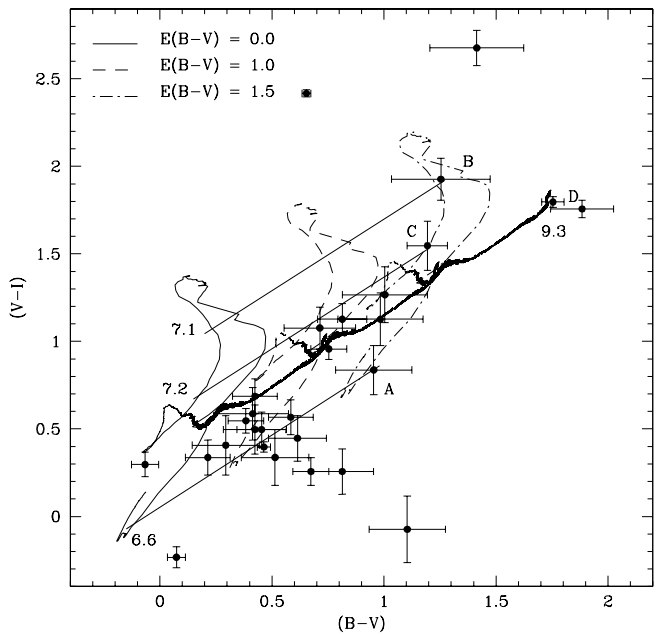,width=0.5\hdsize}
}
\caption{Colour-colour diagram of the clusters in the NGC 6240 main
body.  Superimposed on their distribution are the STARBURST99
evolutionary tracks at $M_{\rm cluster}$ = 10$^5$ M$_{\odot}$ reddened
by E$(B-V)$ = 0.0, 1.0 and 1.5 mag, respectively.  The reddening law of
Calzetti (2001) is applied to the theoretical track, but the clusters
were only corrected for the Galactic reddening towards NGC 6240 [0.076
mag in E$(B-V)$]. Labels A, B, C and D indicate ages of 3, 12.6, 15.8 
Myr and 2 Gyr, respectively.} 
\end{figure}

The panels on the left compare the inner and outer annuli star
clusters; the panels on the right compare 
the outer annulus clusters with the
clusters in the galactic tails.  We discussed (Sect.  4.1)
how the interplay between age and reddening in the two-colour diagram
makes the age determination very poor; and this is reflected in the top
panel of Figure 13, where the clusters in the outer annulus tend to be
younger ($3 - 4$ Myr) than the inner annulus. 
The mass probability function of the inner annulus flattens at 
(middle panel of Figure 13)
about 10$^7$ M$_{\odot}$, but this does not necessarily 
represent a slope break. Indeed, the
combination of sample completeness and photometric errors
affects more severely the probability function at lower
masses (cf. Figure 10). 

Similarly, confirmation of the reality of the apparent peak
near 10$^{5.5}$ M$_{\odot}$ in the outer annulus also
requires deeper, multiwavelength observations which will allow
better age and mass determinations.
The outer annulus is also less reddened (0.2 $\le$ E$(B-V) \le$ 0.8 mag)
than the inner, whose reddening distribution shows a non-negligible tail
at E$(B-V) \ge$ 1 mag.  The plots on the right in Figure 13 show that
the clusters in the outer annulus and in the galactic tails 
statistically are indistinguishable 
in terms of their mass and reddening distributions, with the
latter perhaps being a few Myr older. 

\subsection{Comparisons with other interacting galaxies}

Similar cluster ages and masses have been found in a number of galaxy
mergers, though for this comparison a caveat has to be taken into account, that selection
effects may limit our sample to the brighter and younger (giving the photometric
bands used) clusters in NGC 6240.
\par\noindent
From the point of view of morphology, NGC 6240 and NGC 3256 are quite similar:
they both have extended tails, double nuclei and a FIR luminosity of about
3 $\times$ 10$^{11}$ L$_{\odot}$. The HST observations of Zepf et al. (1999)
have revealed that NGC 3256 has about 1000 young clusters, characterized by
a broad range of colours with little or no correlation between colours and
luminosity. This finding suggests that either low-mass clusters are destroyed
over time or have ages younger than 20 Myr; this second hypothesis is also
observed among the clusters in NGC 6240.

In the case of the Antennae, Whitmore et al.  (1999) derived cluster
ages that depend on location within the galaxies: the youngest clusters
($<$ 5 $\times$ 10$^6$ yr) are preferentially located at the edge of the
dust overlap region, clusters with ages between 5 and 10 Myr are
especially found in the Western Loop, 100 Myr-old clusters are placed in
the NE star-forming region and a population of clusters at an age of 500
Myr seems to be more concentrated in the NW extension of the Antennae. 

\begin{figure*}
\centerline{
\psfig{figure=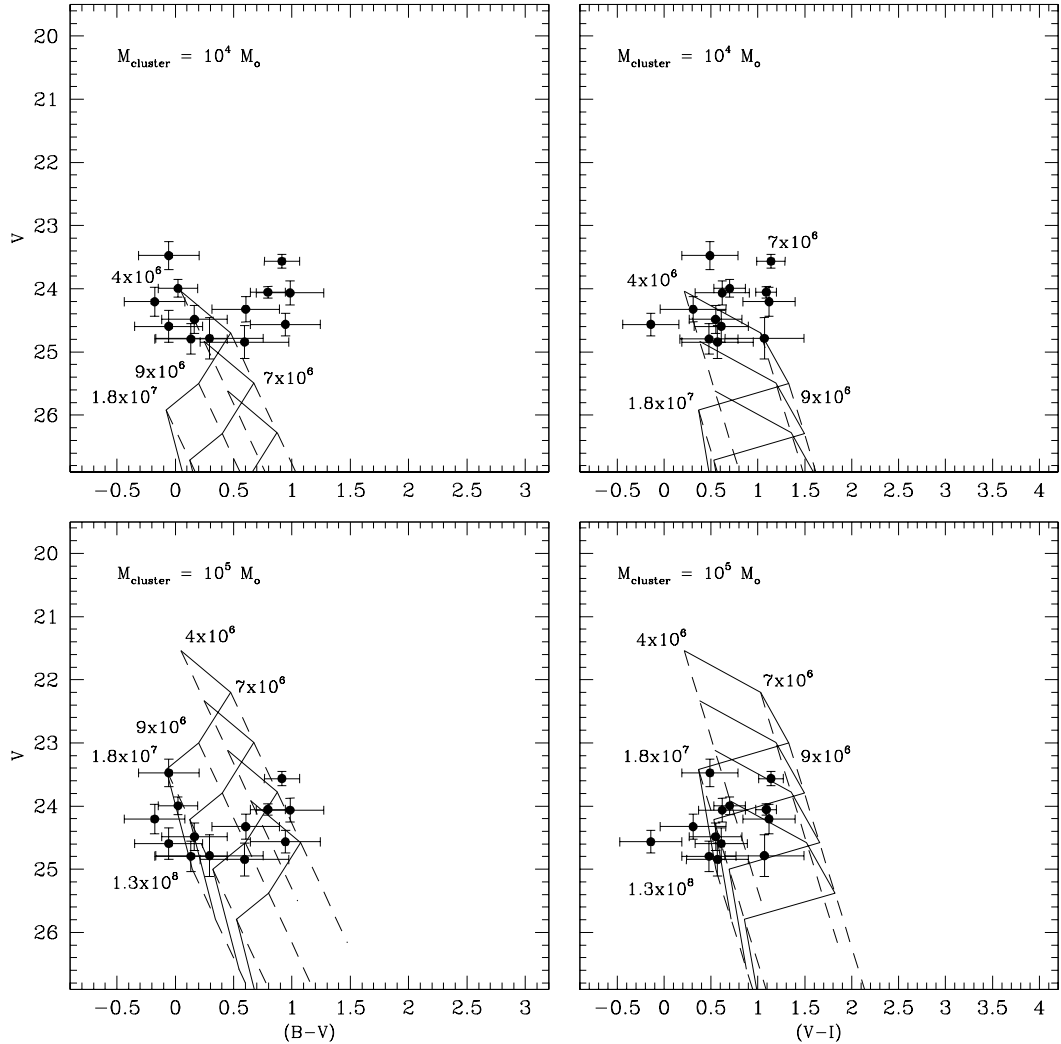,width=0.8\hdsize}
}
\caption{$V$ vs. $(B-V)$ and $(V-I)$ colour-magnitude diagrams for the clusters
identified in the NGC 6240 tails.  They are corrected for the Galactic
extinction towards NGC 6240, E$(B-V)$ = 0.076.  STARBURST99 evolutionary
tracks, computed for an LMC metallicity and two different cluster
masses, are represented by their more significant time mesh points.  At
each given mass, the model is reddened following Calzetti's (2001)
reddening law.  The adopted intrinsic E$(B-V)$ increases from 0.  to 1.0
mag in 0.2 mag steps.}
\end{figure*}

Comparison to the dynamical simulations of Mihos \& Hernquist (1994)
indicates that the 500 Myr-old clusters are probably the relics of the
Antennae's first encounter, when the tidal tails were ejected.  After
their first encounter, the two galaxies separated, only to merge later
in the current configuration and form the younger clusters currently
observed.  An age of 500 Myr intersects the age distributions in Figure
9 at N$_{\rm bin}/$N$_{\rm tot}$ = 60 per cent, where the number of
clusters in NGC 6240 that are statistically likely to 
be older than 500 Myr drops to 20 per cent.  
The majority of the clusters in NGC 6240 seem to have younger ages,
perhaps pointing at a more recent first encounter between the two parent
galaxies. Their estimated masses span a similar range to that observed
for the clusters in the Antennae (Whitmore et al. 1999). 

Other examples of possibly merging galaxies are NGC 7252 and NGC 7673,
where large populations of stellar clusters have been identified
using HST/WFPC2 observations.  The clusters in NGC 7252 split into
three groups: the SSCs in the galactic inner disk are younger
than 10 Myr; the clusters outside the disk with ages of $\simeq$ 750
Myr; and the old globular clusters, relics of the two parent galaxies
(Miller et al.  1997).  In particular, the number of 10 Myr-old clusters
is equivalent to 70 per cent of the total sample of 750 Myr-old clusters. The
mass of the young clusters is estimated to be greater than 10$^5$
M$_{\odot}$, and up to $10^7 - 10^8$ M$_{\odot}$.  In particular, the
brightest cluster in NGC 7252, W3 (Whitmore et al.  1993), is
characterised by an age of about 300 Myr and a mass of about 3.7
$\times$ 10$^7$ (Maraston et al.  2001).  An age of $\ge$7 $\times$ 10$^8$
yr is applicable to less than 50 per cent of the clusters in NGC 6240,
according to Figure 9, while an age younger than 10 Myr is met by the
majority of the cluster sample in NGC 6240.  It also appears that the
more luminous clusters in NGC 6240 and NGC 7252 are equally massive. 
A further example is
presented by NGC 3597, which is a merging system 
resembling NGC~6240 with a double nucleus
and a population of SSCs younger than 10 Myr (Forbes \& Hau
2000). 

\begin{figure*}
\centerline{
\psfig{figure=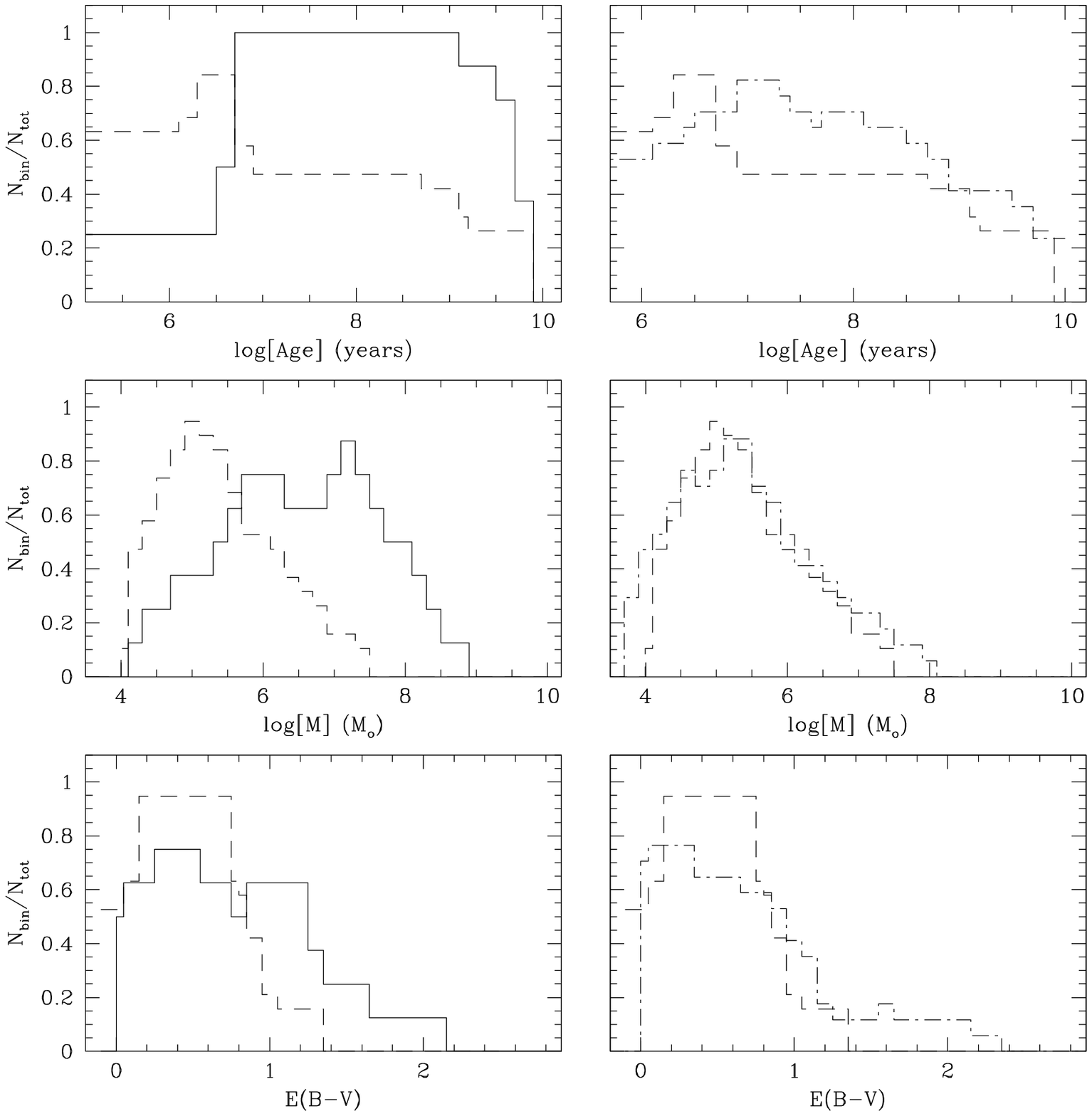,width=0.8\hdsize}
}
\caption{As in Figure 9, but for the outer annulus (dashed line), inner
annulus (solid line) and the cluster sample in the NGC 6240 tails
(dot-dashed line).}
\end{figure*}

The SSC population in the smaller starburst galaxy NGC 7673 is
biased towards young ages: a large number of clusters are dated at
$\simeq$ 6 Myr, and a smaller fraction of clusters shows 
ages between 6 and 20 Myr (Homeier et al. 2002). The youngest SSCs are
found mostly in the gaseous clumps of the galaxy.  The cluster
mass ranges between 0.5 and 5 $\times$ 10$^4$ M$_{\odot}$, and it rises
up to 3 $\times$ 10$^5$ M$_{\odot}$ for the cluster dominating clump F. 
It is interesting to notice that no luminous cluster in NGC 7673 has
been found with an age of $>>$ 10$^8$ yr, like those seen in the Antennae
and NGC 7252. These intermediate age star clusters 
usually are considered to mark the epoch of the parent
galaxies' first encounter (Mihos \& Hernquist 1994). With respect to
the lack of clusters older than 10$^8$ yrs, 
NGC 7673 is very similar to NGC 6240; this suggests that the 
presence of star clusters associated with the first encounter 
is not connected to the scale of the interaction. 

SSCs have also been studied in interacting pairs of galaxies,
such as NGC 1741 which belongs to the Hickson Compact Group HCG 31
(Hickson et al.  1989).  Johnson et al.  (1999) find an age of
$\simeq$ 4 Myr for the majority of the clusters and dated a smaller
number of star clusters to be as old as $\simeq$ 100 Myr.  
The average cluster mass
is about 3 $\times$ 10$^4$ M$_{\odot}$.  Compared to NGC 6240, the
clusters in NGC 1741 are definitely younger and less massive. 

M82 provides another example of a starburst induced by an interaction, 
albeit a very mild one, with its neighbour, the M81 giant spiral system. 
The study by de Grijs et al. (2001, 2003) shows that compact star clusters in 
M82 have ages extending from very young SSCs ($<$10~Myr; e.g., 
Satyapal et al. 1997) 
to a substantial cluster population that formed 0.6$-$1~Gyr in the past. 
The median mass of the M82 clusters is estimated to be $\sim$10$^5$~ 
M$_{\odot}$ (de Grijs et al. 2001, 2003),
and the most massive star clusters have about 10 times the median 
mass (Smith \& Gallagher 2001). Thus M82 appears to have fewer massive 
clusters than more luminous starbursts, such as the Antennae, NGC~6240, or
NGC~7252. 
However, M82 resembles these larger 
starbursts in that the young SSCs are centrally located, an effect that 
is not pronounced in all starbursts (e.g., NGC~7673).

In conclusion, we note that recent and 
ongoing mergers, i.e.  the Antennae, NGC
6240, and NGC 7252 are characterised by prolonged starburst episodes, 
which produce star clusters during three distinct epochs:
$\simeq$ 10 Myr, a few $\times 10^8$ yr, and the ancient globulars, all 
having masses between 10$^4$ and 10$^6$ M$_{\odot}$.  
However, this behaviour is not universal. NGC~1741, an  
interacting galaxy in a compact group, for example, seems to have 
experienced a more ``episodic'' and ``modest'' star formation, 
at least in terms of star cluster formation. 
A large majority of its luminous star clusters are  
SSCs with ages $\simeq$ 4 Myr that are biased towards masses as low as 10$^4$
M$_{\odot}$. M82 then sits in the middle, as another case where cluster 
formation appears to have been episodic and to not widely extend to the highest 
masses. 

NGC 3921 completes the above picture for mergers. It has already
evolved into a 0.7 Gyr-old proto-elliptical (Schweizer et al.
1996).  Its main star cluster population dates as old as $250 - 750$ Myr
(depending on the adopted metallicity) and apparently does not include
clusters as young as $\simeq$ 10~ Myr.
Based on the cluster luminosity function in NGC 3921,
Schweizer et al.  (1996) argued that one of the parent galaxies of NGC
3921 must have been relatively
gas poor compared to NGC 7252, which hosts clusters
of similar age.  This could also have limited the star formation so that
clusters were born only during a limited phase of the interaction. 
On the other hand, in older mergers a variety of dynamical processes
could lead to the disruption ofof  less-bound stellar clusters, so we may
not be seeing the entire picture (cf.  Chernoff \& Weinberg 1989).

\section{Summary}

We have investigated the star clusters in HST/WFPC2 archival 
images of NGC 6240.  
Using a statistical approach, we
fit the observed colours to STARBURST99 stellar population 
models to estimate representative 
cluster ages, masses and intrinsic reddening values.  Our fitting
technique, which assumes for NGC 6240 a LMC-like metallicity, relies 
on observed colours corrected for the Galactic
extinction in the direction of NGC 6240 and develops through five steps:
{\it (i)} the synthetic colours at each time mesh-point of the model are
reddened by E$(B-V)$ increasing from 0.  to 5.0 mag in steps of 0.01
mag; {\it (ii)} at each step in E$(B-V)$, a $\chi_{\rm tot}^2$ is
computed as the squared sum of the colour differences between
observations and evolutionary tracks weighted by the observational
errors; {\it (iii)} for each time mesh-point of the model, a minimum
$\chi_{\rm tot}^2$ is extracted from all the $\chi_{\rm tot}^2$ values
obtained by varying the reddening, together with its corresponding
E$(B-V)$ and model age, magnitudes and colours; {\it (iv)} once all the
time mesh-points of the model have been fitted as above, the smallest
$\chi_{\rm tot}^2$ is identified among all the $\chi_{\rm tot}^2$ minima
and its corresponding E$(B-V)$, model age, magnitudes and colours are
finally assigned to the given cluster; {\it (v)} the selected model
magnitudes are reddened by the final selected E$(B-V)$, corrected for
the distance modulus to NGC 6240 and scaled to the observed apparent
magnitudes. 

The cluster mass estimate 
is computed by multiplying the mass of the
adopted evolutionary track by a visual 
magnitude scaling factor. 
Unfortunately, only two colours, $(B-V)$ and $(V-I)$, are available for
the clusters to determine two independent parameters, age and reddening. 
Therefore, our fitting procedure does not yield unique solutions for 
any individual clusters. Instead we compute a range of
solutions, all of whose minimum $\chi_{\rm tot}^2$ values are less than 
unity, and which can be considered equally consistent with the data. 
The solution for each cluster finally 
is represented by a statistically likely range of ages, an interval
of reddening values and a range of masses.

For each parameter, age, reddening and mass, the cluster solutions
at $\chi_{\rm tot}^2 \le$ 1 were combined into histograms, normalised by
the total number of sample clusters. This is equivalent to a
probability distribution, i.e.  the histograms give the
probability for the clusters to have
a specific age, mass or reddening.  
The caveat here is that this statistical approach, by taking
into account the large ranges of values consistent with the data,
smears the true mass distribution of
the clusters and produces a peak at the most probable value of
the cluster mass.
Based on these probability distributions, we find:
\par\noindent
(1) While the age range spanned by the luminous 
star clusters in NGC 6240 may be quite
large (from 2 $\times$ 10$^6$ to 10$^9$ yr), the more probable cluster 
ages lie between 5 and 13 Myr. 
\par\noindent
(2) The most probable mass 
for the observed luminous clusters is $(1 - 2) \times$ 10$^5$ M$_{\odot}$;
masses are distributed between 3 $\times$ 10$^4$ and 10$^6$ M$_{\odot}$,
with a small fraction of clusters possibly being as massive as 10$^8$
M$_{\odot}$. 
\par\noindent
(3) For the majority of the observed 
clusters, E$(B-V)$ varies between 0.  and 0.8
mag, with a small fraction of observed 
clusters (5 per cent) suffering from 1.3
$\le$ E$(B-V) \le$ 2.4 mag. More examples of heavily obscured clusters may 
exist, but would not be seen in optical images. The outer clusters 
tend to be less reddened than those near the center of the system.
\par\noindent
(4) The star clusters which statistically are most 
likely to have masses exceeding 10$^6$ M$_{\odot}$ are centrally 
concentrated in NGC~6240. The possibility exists for a slope break 
in the cluster mass function at a higher mass than in the Antennae
(Whitmore et al. 1999). 
However, selection effects and dust obscuration
can mimic the presence of a slope break. In the case of the
outer clusters, for which extinction and age effects are less severe,
the distribution of cluster masses extends as a 
power law up to $\sim$ 10$^{5}$ M$_{\odot}$. Clearly, deeper observations
over a wider wavelength range could reveal whether these are true
peaks in the mass distribution of the clusters in NGC 6240. 

In terms of their distributions in  mass and age, 
the clusters in NGC 6240 closely resemble
those in the Antennae and in NGC 7252.  There is also evidence that
some of the 
compact knots of star formation in warm ULIG galaxies can have a mass of
10$^6 - 10^8$ M$_{\odot}$ -- 
as observed in NGC 6240 (cf. Scoville et al. 2000). 
Similar age distributions of 
clusters are found in other starburst/merging galaxies such as 
M82 or NGC 7673, where clusters 
are less massive, about $(0.5 - 10) \times$ 10$^4$
M$_{\odot}$.  Cluster masses may be even lower in some interacting
pairs of galaxies, i.e.  3 $\times$ 10$^4$ M$_{\odot}$ for the 4 Myr-old
SSCs in NGC 1741. 

Therefore, a plausible scenario is that
only the more massive merging systems, such as 
those producing LIGs and ULIGs, can form very high 
mass compact clusters (M $\ge$ 10$^7$~M$_{\odot}$, cf. NGC 7252 Maraston et
al. 2001, NGC 6745 de Grijs et al. 2003), 
while less massive or gas-poor merging/interacting galaxies favour the 
formation of lower mass clusters. Since examples of extremely massive 
star clusters are not seen in globular cluster systems, this type of 
compact system either was not formed in the past, or tends to not to survive
for more than a few Gyr. These considerations may not be inconsistent
with a cluster initial mass function following a power law. Only observations
of star clusters on a large wavelength baseline can remove the age - reddening
degeneracy, allowing an accurate determination of cluster ages, masses and
hence the cluster initial mass function.


\section{Acknowledgments} We thank the anonymous referee for valuable comments
that helped to improve the paper. 
We also thank F. van den Bosch and 
M. Carollo for valuable discussions.
RdeG acknowledges partial financial support
from ST-ECF during an extended visit when this project was initiated, 
and JSG is pleased to acknowledge support from the University of 
Wisconsin Graduate School and from ESO for a short term visit to its 
Garching Headquarters where this work was initiated. 
This paper is based on observations made with the NASA/ESA Hubble Space
Telescope, obtained from the data archive at the Space Telescope
Institute.  STScI is operated by the association of Universities for
Research in Astronomy, Inc.  under the NASA contract NAS 5-26555.  This
research has made extensive use of NASA's Astrophysics Data System
Abstract Service.

\end{document}